\documentclass[11pt]{revtex4}
\usepackage{amsmath,amssymb}
\usepackage{epsfig}
\begin{document}
\title{Separability in 1+1 Dimensions in Classical  Nonlinear Fields}
\author{Azizollah Azizi and Mohammad Mohammadi} \address{Physics department, College of Sciences, Shiraz University, Shiraz 71454,
Iran} \email{azizi@physics.susc.ac.ir}
\date{\today}
%
\begin{abstract}
Solitary wave and soliton solutions of nonlinear equations are well known for physicists. A soliton is a solitary wave with some outstanding features which make it reasonable to be studied seriously in nonlinear systems. In fact most of the nonlinear systems which have solitary wave solutions, has no soliton solutions. To realize a solitary wave as a soliton, we must prepare some initial conditions to collide two or more solitary wave solutions. In fact it is not possible to prepare such initial conditions for any nonlinear system with solitary wave solutions. In this paper we study the conditions that a system should have, to prepare a combination of its single solitary wave solutions as an initial condition for collision. These systems accept a combination of separated single solitary waves as an initial condition, so we call them separable systems. We see a system with periodic potential that zero is one of its vacuum points is separable. We observed that separable systems have two distinct set of solitons, but in general, if we collide members of different sets, they may not behave like solitons.
\end{abstract}
\maketitle
\section{Introduction}
Solitary wave and soliton solutions of nonlinear systems in classical field theories are very interested in physics. Normally people use the word solitary solution in its own purpose and instead of a soliton solution. A solitary wave is a specific solution, that moves without deforming and loosing its energy, ie. without dispersion and dissipation \cite{rajarama}. A soliton solution is such a solitary wave which retain its shape in space after collision with other solitary waves \cite{rajarama,coleman,Drazin,Scott}.
\par
The word collision is important to justify a soliton. A question arise here; dose it possible to collide two or more solitary wave solutions with each other?  For a large group of non-linear systems, there is no way to collide any arbitrary number of its single solitary wave solutions. In fact, in these systems, there is no way to construct a combination of single solitary solutions as an initial condition. A successful system which is always used to explain solitons is the sine-Gordon system in 1+1 dimension, that any combination of its single solitary wave solutions is possible \cite{rajarama,Scott}. But in $\varphi^{4}$ system, we can only combine the solitary solutions in a specific way with some treatments, to construct an initial condition for collision \cite{phi4,phi42,phi43,phi44}.
\par
We search for those systems, that we could combine any set of their solitary wave solutions, to construct an initial condition for collision. If we are able to prepare such an initial condition at far past, we may let the equation of motion to evolve this prepared initial condition as the time passes. This is the meaning of the collision. We  name such system, ``\emph{separable}'' system. A separable system is a system which has no restriction, to make a combination of any arbitrary solitary wave solutions as an initial condition to study the collision between solitary wave solutions (such as sine-Gordon system). Indeed the separable does not mean that the prepared initial condition will necessarily separate enough time after the collision. In fact the initially combined solitary waves may separate after an enough long time (after they collided), or may dissipate or change after the collision.
\par
We will bring a definition for the separable concept. We study our cases in 1+1 dimensions. We find out, that a system with a periodic potential, which has the zero as one of its vacuum points, is separable. Based on these conditions one may build a separable system from a non-separable one, such as $\varphi^{4}$ system. At the end, we study the collision between a large number of kinks and anti-kinks of different separable systems. In general, in a separable system a combination of the kinks (only the kinks) in a collision process behaves like solitons. This result is true for a combination of the anti-kinks (only anti-kinks) as well. But, if we combine kinks and anti-kinks, they will not behave such as solitons, ie. they will destroy or change after collision. So, we conclude that in a separable system, except in the sine-Gordon system, there are two soliton set solutions which are named ``\emph{kink set}'' and ``\emph{anti-kink set}''. The sine-Gordon system has only one soliton set.
 \section{separable  systems }
We restrict ourselves to those systems in 1+1  dimensions with a general lagrangian density  as below
 \begin{equation} \label{lag2}
    \pounds=\frac{1}{2}\\\dot{\varphi}^{2}-\frac{1}{2}\\\varphi'^{2}-U(\varphi).
 \end{equation}
where prime and dot represent space and time derivatives respectively, $\varphi$ is a scalar field and the potential $U(\varphi)$ is a function of space and time via the field $\varphi$. The equation of motion and the energy (Hamiltonian) density correspond to (\ref{lag2}) are \cite{mandl}
\begin{eqnarray}
   \ddot{\varphi} - \varphi'' &=& -\frac{dU(\varphi)}{d\varphi}, \label{eqm}\\
   \varepsilon(x,t) &=& \frac{1}{2}{(\dot{\varphi})}^2 + \frac{1}{2}(\varphi)^2+U(\varphi)\label{ham}.
\end{eqnarray}
For nonlinear field systems, the equation of motion (\ref{eqm}) may have some kind of solutions, so called solitary waves.  A solitary wave is a specific solution of nonlinear systems which its energy density is localized in space and time, and moves with a constant velocity. Solitary waves of those systems with Lagrangian (\ref{lag2}), are called kink solutions with positive topological charges, and anti-kink solutions with negative topological charges \cite{rajarama}. Normally, people characterize kinks by its sectors, ie. $\{\varphi(-\infty)=\varphi_{1},\varphi(\infty)=\varphi_{2}\}$, and anti-kinks by $\{\varphi(-\infty)=\varphi_{2},\varphi(\infty)=\varphi_{1}\}$, which $\varphi_{1}$ and $\varphi_{2}$ are each two next vacuum points, ie. the minimum points of potential $U(\varphi)$ \cite{rajarama}.
\par
People who like to collide solitary waves, normally consider an initial condition that at the far past, consists of some separate single solitary waves with arbitrary initial velocities and positions. The energy density, $\varepsilon(x,t)$, of this solution should have the form
 \begin{equation}\label{add of M Energy-density1}
     \varepsilon (x,t)=\sum_{i=1}^N {\varepsilon_{i}(x-v_{i}t-x_{i})}.\quad (t=-\infty )
\end{equation}
 where $\varepsilon_{i}(x-v_{i}t-x_{i})$ is the energy density of a single solitary wave solution that moves with velocity $v_{i}$ and its initial position at $t=0$ is  $x_{i}$ \cite{rajarama,phi4,dsg1,dsg2,dsg3,dsg4,dsg5}. $N$  represents the number of single solitary wave solutions which are added together. For sine-Gordon system, with a famous potential $U(\varphi)=1-\cos\varphi$, we can simply add some arbitrary number of single solitary wave solutions as below
 \begin{equation}\label{add of M Solitary-Wave1}
       \phi(x,t)=\sum_{i=1}^N {\varphi_{i}(x-v_{i}t-x_{i})},\quad (t=-\infty )
\end{equation}
in order to achieve a combination of $N$ distinct energy density lumps like relation (\ref{add of M Energy-density1}). But in general, we cannot do the same with single solitary wave solutions of any nonlinear system, which will be explained later in this paper. We may name a system, in which a sum over some of its distant arbitrary single solitary waves at far past, such as (\ref{add of M Solitary-Wave1}), makes an energy density of the same number of single separate distant lumps in space, like (\ref{add of M Energy-density1}), ``\emph{separable system}''.
\par
 To deduce the definition of a separable system, we may suppose the addition of $N$ arbitrary distant solitary-wave solutions with different velocities such as relation (\ref{add of M Solitary-Wave1}). It is straightforward to obtain the energy-density of this field (ie, (\ref{add of M Solitary-Wave1})) directly, from the relation (\ref{ham})
\begin{equation} \label{energy}
     \varepsilon'(x,t)=\frac{1}{2}{(\dot{\phi})}^2+\frac{1}{2}(\phi')^2+U(\phi).
\end{equation}
 In another way, we may sum over single energy densities of single solitary wave solutions as we did in (\ref{add of M Energy-density1}). If $\varepsilon'(x,t) = \varepsilon(x,t)$ for every arbitrary number $N$ of single solitary waves, we say this system is separable.
  If a system is separable, we will be able to collide each arbitrary number of its single solitary wave solutions. Just a comment is required here; Some systems are partially separable, ie. we may construct a combination of some single solitary wave solutions (not any arbitrary number) of the system, which are separable. We explain this comment later.
 \section{separability conditions}
 Sine-Gordon system is  the simplest example of those systems which are  separable. What conditions does sine-Gordon system have, that make it separable? Before we answer to this question, we study two examples of nonseparable and partially-separable systems to attain a good background.
\par
\begin{figure}
   \epsfxsize 16cm
   \centering{\epsfbox{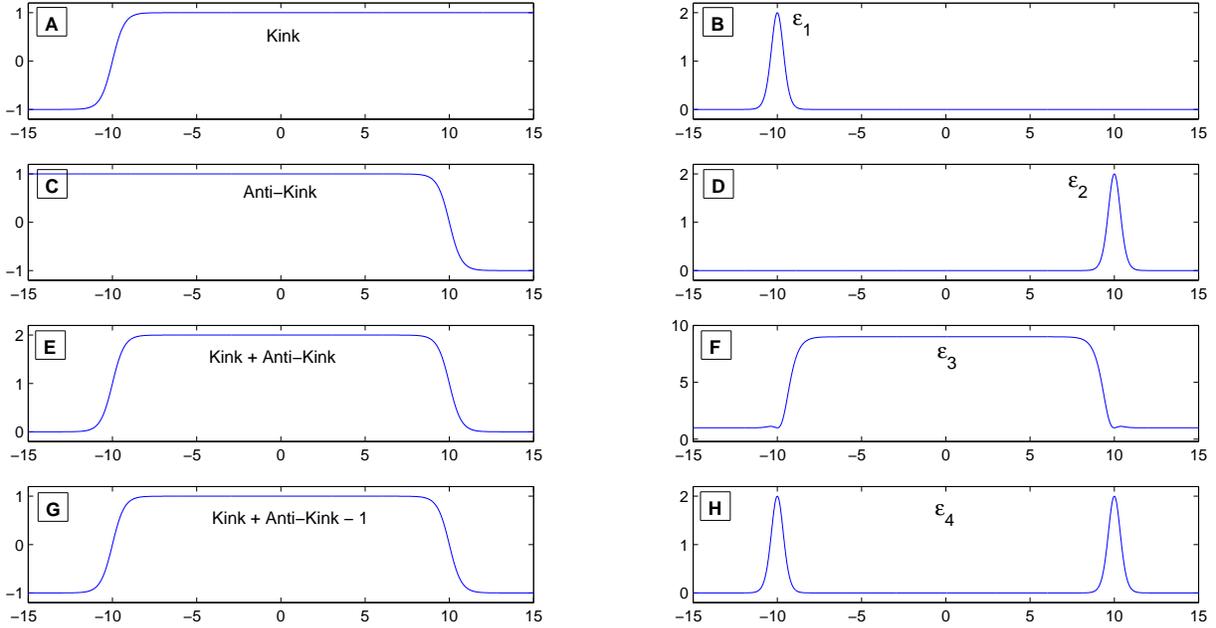}}
   \caption{Column 1 shows fields, while column 2 shows energy densities. First row is a kink, second row is an anti-kink, third row is the combination of a kink and an anti-kink just by adding them where they are far apart, and the forth row is given by modifying the third row, in $\varphi^{4}$ system.   \label{qq}}
\end{figure}

 As an example, we study the $\varphi^{4}$ system with the potential
\begin{equation}\label{pot phi4}
    U(\varphi)=(\varphi^{2}-1)^{2}.
\end{equation}
 This system has two vacuum points at $\varphi=\pm1$, so it has infinite solitary-wave solutions with different velocities that is being classified with $\{-1,1\}$ and $\{1,-1\}$ sectors, which can be distinguished in kink set with positive topological charges and anti-kink set with negative topological charges as well. Figs.~\ref{qq}-A and C show a kink and anti-kink fields, and Figs.~\ref{qq}-B and D show their energy densities respectively in $\varphi^{4}$ system.
 \par
 If in the above system we combine (add) a kink when it is centered at far left and an anti-kink when it is centered at far right at $t=-\infty$ (Fig.~\ref{qq}-E), we receive the energy density in Fig.~\ref{qq}-F, while we were interested in Fig~\ref{qq}-H. We may add $\varphi=-1$ to the combination of a kink and an anti-kink at far past as below (Fig.~\ref{qq}-G)
 \begin{equation}\label{add}
       \phi(x,t)=\varphi_{\textrm{k}}(x-v_{1}t-x_{1})+\varphi_{\textrm{a-k}}(x-v_{2}t-x_{2})-1.\quad (t=-\infty )
\end{equation}
to arrive at Fig~\ref{qq}-H. In fact when we add the two solutions at far past, when they far apart, the combined field is raising from 0 at $x=-\infty$ to 2 at $x=0$ and then to 0 at $x=+\infty$. The potential $U(\varphi)$ is not vanishing at both $x=\pm\infty$ and $x=0$, so the energy of the combined field is not finite as we can see from Fig.~\ref{qq}-F as well.
\par
When we add -1 to the combined (kink and anti-kink) field, we shift it such that the field lays between -1 and +1 at whole space, ie.~it lays in a sector. This yields the potential $U(\varphi)$ vanishing at both $x=\pm\infty$ and $x=0$, so the energy of the shifted combination (ie. Eq.~(\ref{add})) becomes finite (Fig.~\ref{qq}-H).
 To build a combination of a kink and an anti-kink, it is important to know which of them is in the left (or right). For example, if we want to construct another combination of  a kink when it is centered at far right and an anti-kink when it is centered at far left at $t=-\infty$, we have to  add +1 instead of -1 to arrive at favorite form (Fig.~\ref{qq}-H).
 \par
\begin{figure}
   \epsfxsize 16cm
   \centering{\epsfbox{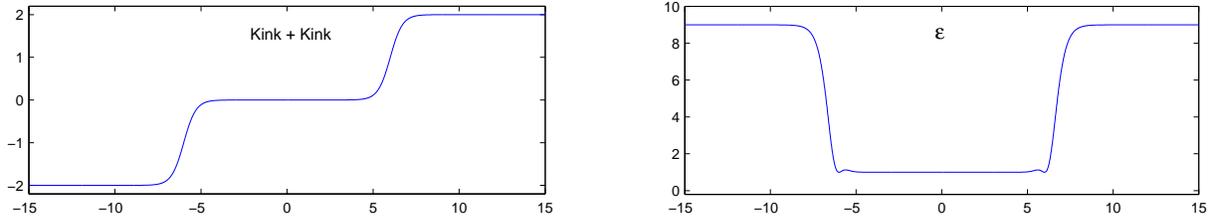}}
   \caption{A combination of two kink (by adding them) and its energy density  where they are far distant, in the $\varphi^{4}$ system.   \label{ss}}
\end{figure}

 If we combine (add) two individual kink solutions, Fig.~\ref{ss}-A, (or two individual anti-kink solutions) when they are far apart at far past, we do not get a suitable energy density, Fig.~\ref{ss}-B. We never can reach to a suitable energy density the same as Fig.~\ref{qq}-H by shifting the combined fields as we did in Eq.~(\ref{add}).
\par
The potential
\begin{equation}\label{ep}
      U(\varphi)=((\varphi-1)^{2}-1)^{2}
\end{equation}
is almost the same as (\ref{pot phi4}), unless the Vacuum points have shifted one unit to the right. If we combine an anti-kink and a kink (anti-kink at left), we get just the right answer, but if we add a kink and an anti-kink (kink at left) we should add +1 to the combination to have a suitable answer.
\par
\begin{figure}
   \epsfxsize 15cm
   \centering{\epsfbox{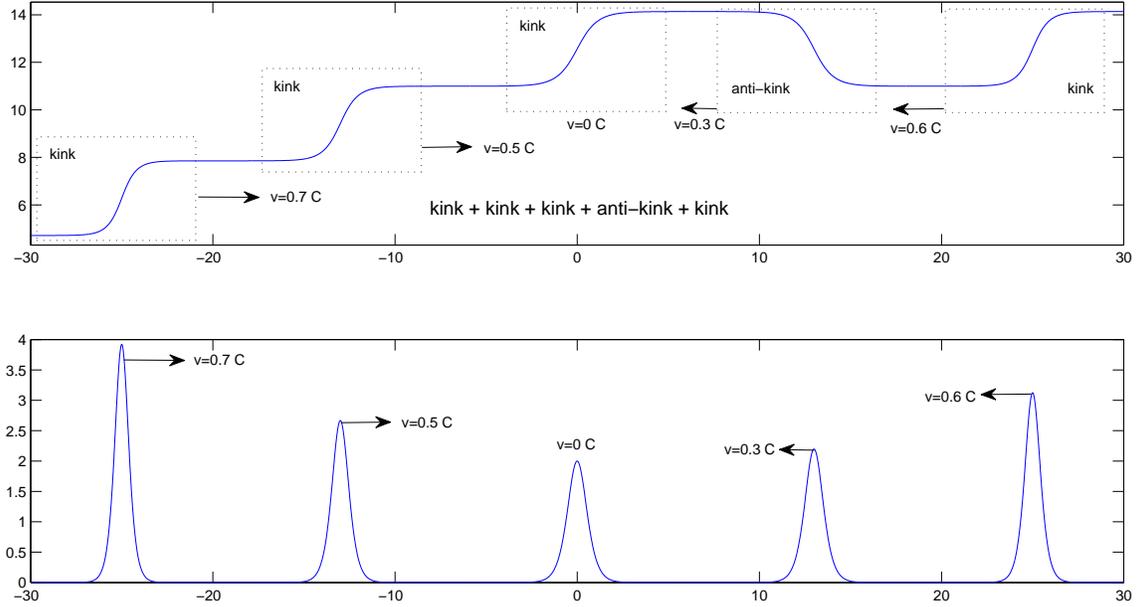}}
   \caption{An arbitrary combination of odd number of solitary waves (four  kink and one anti-kink with arbitrary initial velocities) and its energy density  when they are far at far past,  in $\cos^{2}(\varphi)$ system.   \label{xx}}
\end{figure}
\begin{figure}
   \epsfxsize 15cm
   \centering{\epsfbox{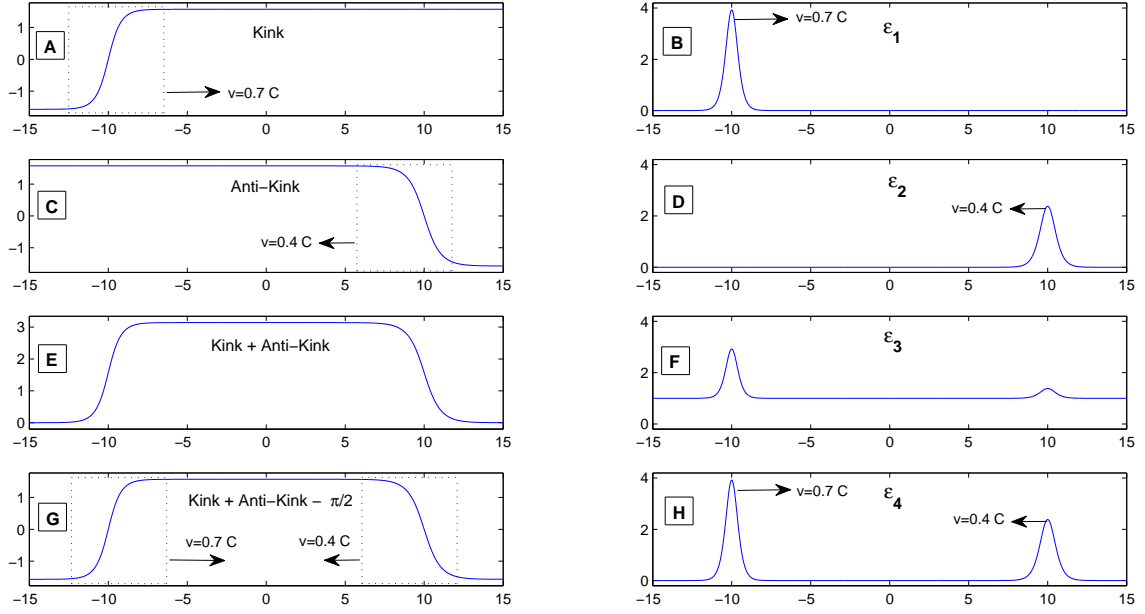}}
   \caption{An arbitrary combination (E) of one  kink (A) and one anti-kink (C) with arbitrary initial velocities (even number of solitary waves),  when they are far distant at far past, results an infinite energy in $\cos^{2}(\varphi)$ system (F). A change in the combination is needed (G) to result a favorite  initial condition (H). Simply, one can see that the Fig. H is exactly the superposition of Figs. B and D, but not the Fig. F. \label{fig2}}
\end{figure}

A system with the potential
\begin{equation}\label{pot cos2}
    U(\varphi)=\cos^{2}(\varphi),
\end{equation}
 is a partially-separable system with vacuum points at $(2n+1)\pi/2$, $n\in\mathbb{Z}$, and infinite sectors. When we combine an odd arbitrary number of solitary wave solutions (kinks and/or anti-kinks), we observe the behavior of a separable system (see FIG.~\ref{xx}). But when we combine an even number of solitary waves (kinks and/or anti-kinks), it behaves not like a separable system (see Fig.~\ref{fig2}-E and F). The reason is that the combination does not lay between two vacuum points, so its related energy becomes infinite.  If we add a constant (a required odd factor of $\pi/2$) to the combination, we achieve a favorite form of a separable solution (see Fig.~\ref{fig2}-G and H).
\par
If we shift the potential (\ref{pot cos2}) by $\pi/2$ in $\varphi$, we will get $\sin^2\varphi$, which is in fact the sine-Gordon potential. So by this simple change, the partially separable system transforms to a full separable system.
\par
Two conditions are required to make a system separable:
 Firstly, the potential must be periodic, and secondly, $\varphi=0$ must be one of the vacuum points. These are the ``\emph{separability conditions}''. Noting that in shifted $\varphi^4$ potential, Eq.~(\ref{ep}), the $\varphi=0$ became a vacuum point, so some combination of single solitary waves, in contrast to the standard $\varphi^4$ system, became separable.
\par
An important subject in the nonlinear systems, is the investigation of solitary waves collision. We always do like to check that the solitary wave solutions, are solitons or not. To check this, we should collide the solitary waves. Here we may ask, does it essentially possible to collide any arbitrary number of solitary solutions? The answer is definitely related to separability conditions.
\section{How can we build a separable  system?}
\begin{figure}
   \epsfxsize 16cm
   \centering{\epsfbox{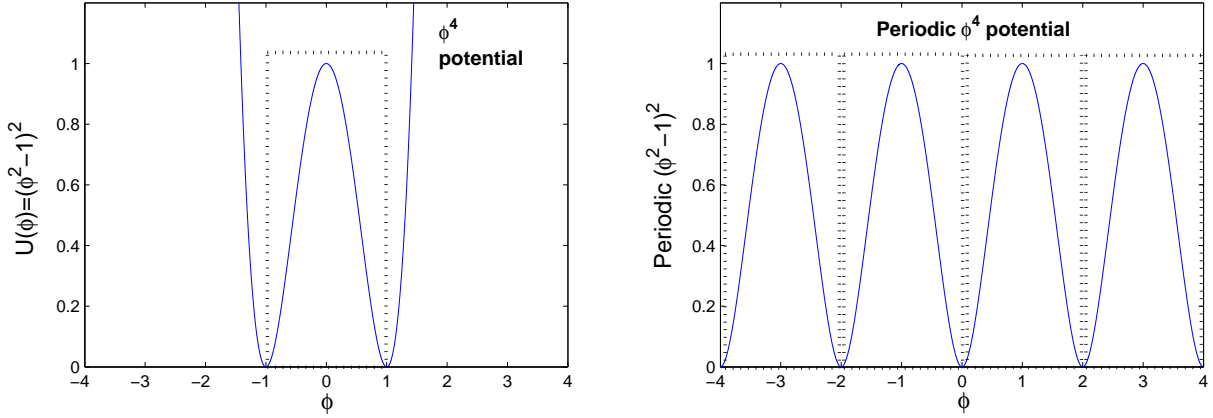}}
    \caption{If we shift the region between vacuum points $\varphi=\pm1$ (inside the dashed rectangular) of $\varphi^{4}$ one unit to right, and repeat if for the other regions, we get periodic $\widetilde{\varphi}^{4}$ system.  \label{aa}}
\end{figure}

As we said before, $\varphi^4$ system is not a separable system, so we cannot collide an arbitrary number of kinks and/or anti-kinks. This system has two vacuum  points at $\varphi=-1$ and $\varphi=1$. The solitary solutions of this system lay between these two vacuum points. So we may use the region between the vacuum points to make a periodic potential. First, we shift the potential as we did in (\ref{ep}), to set the $\varphi=0$ as a vacuum point. Now the solitary waves are laying between $\varphi=0$ and $\varphi=2$. Second, we repeat the region between $\varphi=0$ and $\varphi=2$ to the other regions of the space as we can see in Eq.~(\ref{rp4}) and Fig.~\ref{aa}.
\begin{equation}\label{rp4}
    U(\varphi) =\textrm{Periodic}(\varphi^{2}-1)^{2}= \left\{
    \begin{array}{clc}
         \vdots && \\
        ((\varphi+3)^{2}-1)^{2}, & \quad & -4<\varphi\leq -2,\\
        ((\varphi+1)^{2}-1)^{2}, & \quad & \!\!\!-2<\varphi\leq 0,\\
        ((\varphi-1)^{2}-1)^{2}, & \quad & 0<\varphi\leq 2,\\
        ((\varphi-3)^{2}-1)^{2}, & \quad & 2<\varphi\leq 4,\\
        \vdots& &
    \end{array} \right.
\end{equation}
 We name this new system, ie. the shifted periodic $\varphi^{4}$ system, as $\widetilde{\varphi}^4$ system, which wavy symbol $\sim$ denotes the periodicity of the system.
 \par
We studied the following periodic systems
\begin{eqnarray}
    U(\varphi) &= &\sum_{n=1}^N {\varepsilon_{n}(1-\cos(\varphi))^n}, \quad \quad (\varepsilon_{n}=\textrm{constant})\\
    U(\varphi) &=& \textrm{Periodic}\left(\varphi^4-1\right)^2,\\
    U(\varphi) &=& \textrm{Periodic}\left|\varphi^2-1\right|^3,\\
    U(\varphi) &=& \sin^{2N}(\varphi),\quad \quad (N=1,2,3,\ldots),
\end{eqnarray}
and realized the separability of them.
 \par
 We studied the collision of solitary solutions of the above mentioned periodic systems, by combining various kink and/or anti-kink solutions, when they are far at $t=-\infty$. We observed that, if we let the combination to evolve in time, a combination of only kinks or only anti-kinks, will reappear after collision at $t=\infty$, while a combination of kinks and anti-kinks, will not reappear after collision. We see that the set of kink (anti-kink) solutions, when we regret the anti-kinks (kinks), do have the characteristic of solitons. We may name each set, a ``\emph{soliton set}''. Each of the above periodic systems have two soliton sets. The sine-Gordon system is an exception, which any arbitrary combination of its kink and/or anti-kink solutions will reappear after collision, so, sine-Gordon system has only one soliton set.


\begin{thebibliography}{99}
    \bibitem{rajarama} R. Rajaraman, \textit{Solitons and instantons; an introduction to solitons and
           instantons in quantum field theory}, North-Holland (1989).

    \bibitem{Drazin} P.G. Drazin and R.S. Johnson, \textit{Solitons: an introduction}, Cambridge University
           Press, Combirage (1989).

    \bibitem{Scott} A.C. Scott, F.Y.F. Chiu and D.W. Mclaughlin, \textit{A new concept in applied science}, Proc. I.E.E.E. \textbf{61}, 1443 (1973).

    \bibitem{coleman} S. Coleman, \textit{Classical lumps and their quantum descendants}, 1975 Erice lectures published in ``New phenomena in sub-Nuclear  physics'', Ed. A. Zichichi, Plenum Press, New York (1977).

    \bibitem{phi4} R.H. Goodman and R. Haberman, \textit{Kink-antikink collisions in the $\phi^{4}$ equation: The n-bounce resonance and the separatrix map}, Siam J. Applied Dynamical Systems, Vol. \textbf{4}, No. 4, 1195-1228  (2005).

    \bibitem{phi42}  D.K. Campbell, J.S. Schonfeld and C.A. Wingate, \textit{Resonance structure in kink-antikink interactions in  $\phi^{4}$ theory}, Physica D \textbf{9}, 1–32 (1983).

    \bibitem{phi43}  X. Lu and R. Schmid, \textit{ Symplectic integration of sine-Gordon type systems}, IMACS International Conference Modelling \textbf{98}, Prague (1998).

    \bibitem{phi44} P. Anninous, S. Olivera and R.A. Matzner, \textit{Fractal srtructure in the scalar $\lambda(\varphi^2-1)^2$ theory}, Phys. Rev. D  \textbf{44} (1991).

    \bibitem{mandl}  Mandl and G. Show, \textit{Quantum  field  theory}, Wiley (1993).

    \bibitem{dsg1} M.J. Ablowitz, M.D. Kruskal and J.F. Ladik, \textit{Solitary wave collisions}, Siam J. Appl. Maths \textbf{36}, 428-437 (1979).

    \bibitem{dsg2} R. Ravelo, M. El-Batanouny and C.R. Willis, \textit{Dynamics of kink-kink collisions in the double sine-Gordon system}, Phys. Rev. B \textbf{38} (1988).

    \bibitem{dsg3} D.K. Campbell and M. Peyrard, \textit{Kink-antikink interactions in the double sine-Gordon equation}, Physica D \textbf{19} 165-205 (1986).

    \bibitem{dsg4} V.A. Gani and A.E. Kudryavtsev, \textit{Kink-antikink interactions in the double sine-Gordon equation and the problem of resonance frequencies}, Phys. Rev. E \textbf{60}, 3305-3309 (1999).

    \bibitem{dsg5} M. Peyrard and David K. Campbella, \textit{Kink-antikink interactions in a modified sine-Gordon model} Physica D: Nonlinear Phenomena Volume \textbf{9}, Issues 1-2, Pages 33-51 (1983).
\end{thebibliography}
\end{document}